\documentclass[10pt]{article}
\usepackage[OE]{express}

\begin{document}
\title{Compensation for the setup instability in ptychographic imaging}

\author{Yudong Yao,\authormark{1,2} Cheng Liu,\authormark{1,*} and Jianqiang Zhu\authormark{1}}

\address{\authormark{1}Key Laboratory of High Power Laser and Physics, Shanghai Institute of Optics and Fine Mechanics, Chinese Academy of Sciences, Shanghai 201800, China\\
\authormark{2}University of Chinese Academy of Sciences, Beijing 100049, China}

\email{\authormark{*}cheng.liu@hotmail.co.uk} 



\begin{abstract}
The high-frequency vibration of the imaging system degrades the quality of the reconstruction of ptychography by acting as a low-pass filter on ideal diffraction patterns. In this study, we demonstrate that by subtracting the deliberately blurred diffraction patterns from the recorded patterns and adding the properly amplified subtraction to the original data, the high-frequency components lost by the vibration of the setup can be recovered, and thus the image quality can be distinctly improved. Because no prior knowledge regarding the vibrating properties of the imaging system is needed, the proposed method is general and simple and has applications in several research fields.
\end{abstract}

\ocis{(100.5070) Phase retrieval; (050.1970) Diffractive optics; (120.5050) Phase measurement; (070.0070) Fourier optics and signal processing} 


\section{Introduction}
Coherent diffraction imaging (CDI) is a promising technology for obtaining the complex transmission function of a specimen from the recorded diffraction intensity. As a lens-free technique, CDI can bypass the resolution limits imposed by the poor focusing optics available at short wavelengths \cite{1,2} and can theoretically reach the diffraction-limited resolution. With X-ray and high-energy electrons, a resolution of nanometers or angstroms can be achieved; thus, CDI is becoming an important tool in material and biological sciences \cite{3,4,5,6,7}. Because the performance of traditional CDI algorithms is not very satisfying with regard to convergence, accuracy, and reliability, several improved CDI methods have been proposed \cite{8,9,10,11}. The ptychographic iterative engine (PIE) \cite{12} is a scanning version of the CDI technique where the specimen is scanned through a localized illumination beam to a grid of positions and the resulting diffraction patterns are recorded. Using an iterative scheme with a proper overlapping ratio between two adjacent scanning positions, the modulus and phase of the transmission functions of the specimen and the illumination beam can be reconstructed accurately and rapidly \cite{13,14}. In theory, PIE can easily generate images with a resolution only limited by the numerical aperture of the detector; however, in practice, the resolution is affected by the flaws of the imaging system, especially for imaging with X-ray and electron beams. The coherence of synchrotron radiation sources is not as good as that of a laser beam and has been proven to be the largest barrier for PIE to reach the theoretical diffraction-limited resolution \cite{15,16}. Several algorithms have been investigated to improve the image quality of PIE with partially coherent illumination \cite{17,18,19,20,21,22}. Although these methods require complete or partial prior knowledge of the properties of illumination or time-consuming computation and may slightly compromise the spatial resolution, the quality of the reconstruction can be distinctively improved in many cases. Furthermore, the recently developed technique of Fourier ptychography has potential to avoid the influence of the incoherence for achieving high-quality reconstruction \cite{23,24}. Another factor that limits the practical resolution of PIE is the instability of the imaging system, including the vibration of the mechanical scanning system, the tiny pointing-direction change, and the transverse shifting of the radiation beam \cite{25,26}. Because the wavelengths of the X-ray and the electron beams are much smaller than 1 nm, the tiny departure of the illumination beam from the right position and direction can generate obvious errors in the final reconstruction. Numerous methods have been proposed to efficiently correct the low-frequency vibration with a period far longer than the exposure time of the detector \cite{27,28,29,30}. However, there is no ideal method for effectively handling the high-frequency vibration of the experimental system, which makes the recorded diffraction intensity a summation of many diffraction intensities formed by the changing illumination during the exposure of the detector \cite{31}. Although the influence of the high-frequency vibration can be treated as a type of incoherence of the illumination, the aforementioned methods \cite{17,18,19,20,21,22,31} require complete or partial prior knowledge of the properties of vibration-including the frequencies and amplitude-or massive data processing. However, in practice, the parameters of the vibration of the whole imaging system are difficult to measure in real time. To deal with the image-quality degradation induced by the setup instability, we must examine its influence on the recorded data and the final reconstruction and then develop a method to circumvent this problem.\par
In this study, we mathematically analyze how the recorded diffraction patterns and the final reconstruction are blurred by the instability of the imaging system and then propose a simple numerical method for enhancing the lost high-frequency components.  The proposed method does not require any prior knowledge regarding the characteristics of the high-frequency vibration of the imaging system and can be extended to other CDI methods using X-ray and electron beams to solve the problems related to imaging-system instability.\par
\section{Principle of the method}
\begin{figure}[htbp]
\centering
\includegraphics[width=3in]{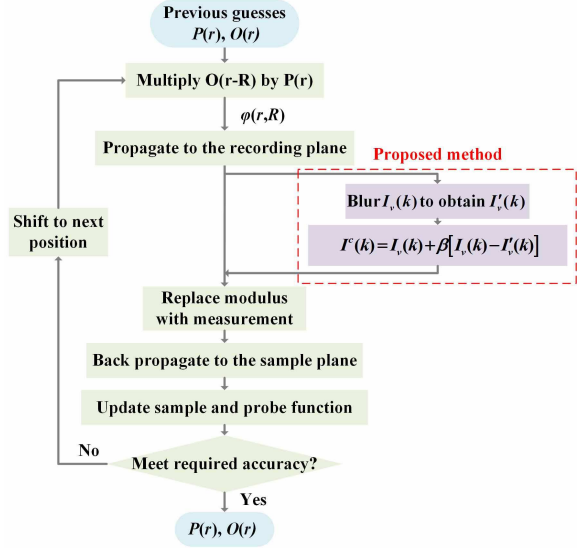}
\caption{Flowchart of the reconstruction process using the conventional PIE method and the proposed method.}
\label{Fig.1}
\end{figure}
In the PIE method, the specimen $O(r)$ is fixed on a two-dimensional (2D) translation stage and illuminated by a localized probe $P(r)$. Assuming that the specimen is sufficiently thin, the exiting wave from the specimen is $\varphi (r,R) = O(r-R)P(r)$, and the recorded diffraction intensity is $I(k,R)\propto|\Im[\varphi(r,R)]|^2$ in most experiments with short-wavelength sources, where $k$ is the reciprocal coordinate with respect to the real space coordinate $r$ in the specimen plane, and $R$ denotes the position of the specimen during the raster scanning. With the recorded diffraction patterns, the complex amplitudes of the specimen and the probe can be reconstructed. The flowchart of the reconstruction process of PIE is shown in Fig. \ref{Fig.1}, and a detailed description is found in the literature \cite{12,13}.\par
\subsection{Influence of the setup instability}
In the PIE algorithm, the illumination probe is assumed to be absolutely static during the data acquisition; however, its direction and position change continually within a small range with respect to the specimen and detector owing to the instability of the imaging system. In most cases, the direction and position of the illumination beam change simultaneously during the exposure of the detector, but for simplicity, we analyze them separately to determine how the recorded data are influenced mathematically and then consider their combined effects. \par
According to the principles of Fourier optics, any illumination beam can be decomposed into a series of spherical waves of different strengths; thus, we can assume a spherical illuminating probe without a loss of generality. The spherical illumination is expressed in Fourier form as $\widetilde{P}(k)=W(k)exp(-j\pi\lambda z k^2 )$, where $\lambda$ is the wavelength of the probe, and $z$ is the distance between the focal spot of the probe and the specimen, and $W(k)$ is determined by the numerical aperture of the illuminating optics. For a short wavelength or in the far field, the diffraction distribution in the detector plane is the Fourier transform of the wave exiting the specimen. The complex amplitude of the diffraction pattern in the detector plane can be expressed as
\begin{eqnarray}
\Phi(k)=\widetilde{P}(k)\otimes\widetilde{O}(k)=\sum_{n}\widetilde{P}(k-k_n)\widetilde{O}(k_n)
\label{eq:one}.
\end{eqnarray}
where $\widetilde{O}(k)$ is the Fourier transform of the transmission function of the specimen. The recorded intensity with static illumination is
\begin{eqnarray}
I(k)&=&|\sum_{n}\widetilde{P}(k-k_n)\widetilde{O}(k_n)|^2\nonumber\\
&=&I_0(k)+\sum_{m\neq n}A_m(k)A_n(k)cos[\theta(k)]
\label{eq:two}.
\end{eqnarray}
The first item $I_0(k)$ is the intensity summation of all diffracted beams, and the second item $\sum_{m\neq n}A_m(k)A_n(k)cos[\theta(k)]$ represents the interference between different spatial-frequency components.
\begin{eqnarray}
&I_0(k)=\sum_{n}|\widetilde{P}(k-k_n)\widetilde{O}(k_n)|^2 \nonumber\\
&A_m(k)=|\widetilde{P}(k-k_m)\widetilde{O}(k_m)|\nonumber\\
&A_n(k)=|\widetilde{P}(k-k_n)\widetilde{O}(k_n)|\nonumber\\
&\theta(k)=\pi\lambda z[2k(k_m-k_n)-({k_m}^2-{k_n}^2)]+\phi_{mn}
\label{eq:three}.
\end{eqnarray}
Here, $k_m-k_n$ is the frequency of the interference fringe formed by the $m^{th}$ and $n^{th}$ diffraction orders, and $\phi_{m,n}$ is the additional phase introduced by the specimen. Considering the coordinate transformation $r_c = \lambda Lk$ on the CCD plane, where $L$ is the distance between the specimen and the CCD, the diffraction pattern intensity can be described in the frequency domain as
\begin{eqnarray}
\widetilde{I}(u)=\widetilde{I}_{0}(u)+\sum_{m\neq n}\widetilde{A}_{m}(u)\otimes\widetilde{A}_{n}(u)|_{u=z(k_{m}-k_{n})/L}
\label{eq:four}.
\end{eqnarray}
where $u$ is the reciprocal coordinate with respect to the real space coordinate $r_c$ in the CCD plane.\par
When considering the pointing instability of the imaging system, the illumination beam tilted an angle of $\alpha$ can be expressed as,
\begin{eqnarray}
P'(r)=P(r)exp(-j2\pi r\frac{sin\alpha}{\lambda})
\label{eq:five}.
\end{eqnarray}
So the Fourier transform of the probe is $\widetilde{P}'(k)=\widetilde{P}(k-\Delta k)$, where $\Delta k=sin\alpha/\lambda$. The recorded intensity with tilted illumination becomes\par
\begin{eqnarray}
I'(k)=|\sum_{n}\widetilde{P'}(k-k_n)\widetilde{O}(k_n)|^2=I(k-\Delta k)
\label{eq:six}.
\end{eqnarray}
Assuming that the vibration in the pointing direction of the illumination beam follows the normal distribution $H_1(\Delta k)=exp(-\Delta k^{2}/K^{2})$, where $K$ is a constant related to the vibrating properties of the imaging system. The recorded intensity with high-frequency vibration in the pointing direction of the illuminating beam can be interpreted as a summation of the diffraction patterns of all possible illuminations with different pointing directions. Thus, it can be expressed as
\begin{eqnarray}
I_{v1}(k)&=&\frac{\int_{-\infty}^{+\infty}{I(k-\Delta k)H_1(\Delta k)d\Delta k}}{\int_{-\infty}^{+\infty} {H_1(\Delta k)d\Delta k}} \nonumber\\
&=&\frac{1}{\sqrt{\pi}K}I(k)\otimes exp(-\frac{k^2}{K^2})
\label{eq:seven}.
\end{eqnarray}
Considering the coordinate transform $r_c = \lambda Lk$ in the detector plane, the recorded intensity can be expressed in the frequency domain as
\begin{eqnarray}
\widetilde{I}_{v1}(u)=\widetilde{I}(u)exp[-(\pi\lambda LK)^2u^2]
\label{eq:eight}.
\end{eqnarray}
Thus, the influence of the vibration in the pointing direction of the illuminating beam acts as a low-pass filter on the ideal diffraction patterns.\par
On the other hand, when the illumination probe suffers from transverse positioning vibration, the recorded diffraction pattern can be expressed as a summation of the diffraction intensities formed by all possible illumination beams with diverse transverse shifts. The probe with a transverse shift of $\delta$ is expressed as $P''(r)=P(r+\delta)$. And its Fourier transform is $\widetilde{P}''(k)=\widetilde{P}(k)exp(j2\pi k\delta)$. The corresponding diffraction-pattern intensity is \par
\begin{eqnarray}
I''(k)&=&|\sum_{n}\widetilde{P}''(k-k_n)\widetilde{O}(k_n)|^2\nonumber\\
&=&I_0(k)+\sum_{m\neq n}A_m(k)A_n(k)cos[\theta(k+\Delta)]
\label{eq:nine}.
\end{eqnarray}
Assume that the transverse shifting of the illuminating beam has a normal distribution $H_2(\Delta)=exp(-\Delta^{2}/D^{2})$, where $D$ is determined by the standard deviation of $\Delta = \delta/\lambda z$. The recorded intensity with high-frequency positioning vibration is a summation of different intensities for different illumination shifts:
\begin{eqnarray}
&&I_{v2}(k)=\frac{\int_{-\infty}^{+\infty}{I''(k)H_2(\Delta)d\Delta}}{\int_{-\infty}^{+\infty} {H_2(\Delta)d\Delta}}=I_0(k)+\\
&&\sum_{m\neq n}A_m(k)A_n(k)cos[\theta(k)]exp[-(\pi\lambda zD)^2(k_m-k_n)^2]\nonumber
\label{eq:ten}.
\end{eqnarray}
The Fourier transform of the recorded diffraction intensity becomes
\begin{eqnarray}
\widetilde{I}_{v2}(u)&=&\widetilde{I}_{0}(u)+exp[-(\pi\lambda LD)^{2}
     u^{2}]\nonumber\\
& &\times\sum_{m\neq n}\widetilde{A}_{m}(u)\otimes\widetilde{A}_{n}(u)|_{u=z(k_{m}-k_{n})/L}
\label{eq:eleven}.
\end{eqnarray}
Compared with Eq. (4), the position vibration degrades the contrast of the interference fringe in the recorded intensity, acting as a low-pass filter.\par
In practice, the pointing and transverse vibration of the illumination beam can occur simultaneously; thus, the Fourier transform of the recorded diffraction intensity is
\begin{eqnarray}
&&\widetilde{I}_{v}(u)=\{\widetilde{I}_{0}(u)+exp[-(\pi\lambda LD)^{2}u^{2}]\nonumber\\
&&\times\sum_{m\neq n}\widetilde{A}_{m}(u)\otimes\widetilde{A}_{n}(u)\}exp[-(\pi\lambda LK)^{2}u^{2}]
\label{eq:twelve}.
\end{eqnarray}
Because $\widetilde{I}_{0}(u)$ has a narrow frequency band for spherical illumination, Eq.(12) can be approximated as
\begin{eqnarray}
\widetilde{I}_{v}(u)\approx \widetilde{I}_{0}(u)+exp(-C^{2}
     u^{2})\sum_{m\neq n}\widetilde{A}_{m}(u)\otimes\widetilde{A}_{n}(u)
\label{eq:thirteen}.
\end{eqnarray}
where $C=\pi\lambda L \sqrt{K^{2}+D^{2}}$. Clearly, the instability of the illumination beam during the exposure of the detector leads to the loss of high-frequency components of the diffraction intensity, and the contrast of the interference fringes is differently suppressed depending on their frequency. As reported in the literature \cite{22}, change in the contrast of the diffraction patterns leads to mathematical ambiguity and generates errors in the final reconstruction.\par
\subsection{Compensation method}
At first glance, it appears that the Fourier component of the ideal diffraction pattern can be obtained by dividing $exp(-C^{2}u^{2})$ to the $\widetilde{I}_{v}(u)$ by each possible $C$ until a satisfactory reconstruction is achieved. However, because this operation can seriously amplify the errors or noise, and the expired resolution improvement can be submerged by them, it cannot be adopted in practice.\par
In the proposed method, the recorded intensities $\widetilde{I}_{v}(k)$ are modified before the conventional PIE process, as shown in Fig.\ref{Fig.1}. The recorded diffraction intensities are deliberately blurred via convolution with the Gaussian function $exp(-k^{2}/K_{v}^{2})$, where $K_{v}$ is a constant chosen to slightly blur the recorded diffraction pattern. The deliberately blurred intensity pattern is expressed in the frequency domain as
\begin{eqnarray}
\widetilde{I'}_{v}(u)&=&\widetilde{I}_{0}(u)+exp(-C^{2}u^{2})exp(-C_{v}^{2}u^{2})\nonumber\\
& &\times\sum_{m\neq n}\widetilde{A}_{m}(u)\otimes\widetilde{A}_{n}(u)
\label{eq:fourteen}.
\end{eqnarray}
where $C_{v}=\pi\lambda LK_v$. We subtract the blurred pattern $\widetilde{I'}_{v}(k)$ from the original recorded diffraction $\widetilde{I}_{v}(k)$ and then add the subtracted pattern to $\widetilde{I}_{v}(k)$ after multiplying it by a constant $\beta$:
\begin{eqnarray}
I^{c}(k)=I_{v}(k)+\beta[I_{v}(k)-I'_{v}(k)]
\label{eq:fifteen}.
\end{eqnarray}
The Fourier transform of the modified intensity pattern is\par
\begin{eqnarray}
\widetilde{I}^{c}(u)&=&\widetilde{I}_{0}(u)+[1+\beta-\beta exp(-C_{v}^{2}u^{2})]\nonumber\\
& &exp(-C^{2}u^{2})\sum_{m\neq n}\widetilde{A}_{m}(u)\otimes\widetilde{A}_{n}(u)
\label{eq:sixteen}.
\end{eqnarray}
Clearly, when $[1+\beta-\beta exp(-C_{v}^{2}u^{2})]exp(-C^{2}u^{2})$ in Eq.(16) is wider than $exp(-C^{2}u^{2})$, $\widetilde{I}^{c}(u)$ is close to the Fourier transform of the vibration-free diffraction pattern, and then the influence of the imaging system instability on the reconstructed image can be remarkably suppressed.
\section{Results}
\subsection{Simulation result}
\begin{figure}[htbp]
\centering
\includegraphics[width=3in]{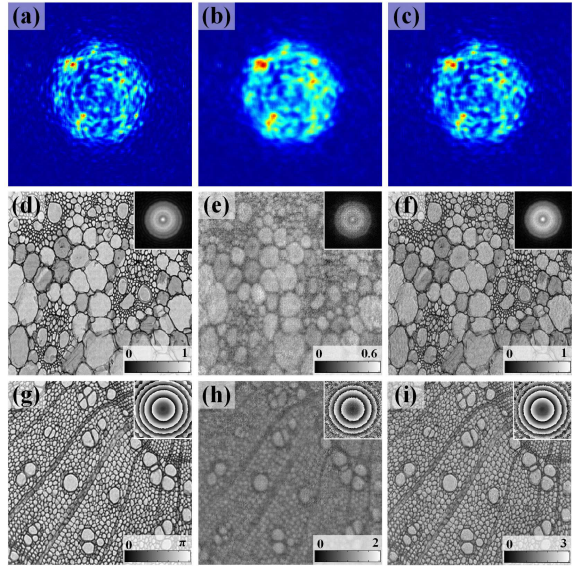}
\caption{Diffraction patterns obtained (a) with stable illumination, (b) with unstable illumination, and (c) by modification via the proposed method; reconstructed modulus obtained (d) with stable illumination, (e) with unstable illumination, and (f) using corrected intensities; reconstructed phase obtained (g) with stable illumination, (h) with unstable illumination, and (i) using corrected intensities. The upper insets in (d)-(i) show the reconstructed illumination fields.}
\label{Fig.2}
\end{figure}
A numerical simulation is performed using the proposed algorithm to check its feasibility. A divergent spherical wave with a wavelength of 632.8 nm is used for the illumination. Two pictures are used as the modulus and phase, respectively, of the specimen. The diameter of the illuminated area on the specimen is 0.74 mm, and $10\times10$ diffraction intensities are recorded with a step size of 0.185 mm. The charge-coupled device (CCD) camera has a resolution of $256\times256$ pixels and a pixel size of 7.4 $\mu$m. For comparison, the diffractive intensities with stable illumination are calculated, one of which is shown in Fig.\ref{Fig.2}(a). When the unstable imaging system results in diffraction pattern vibration with variance of $46\mu m$ in the detector plane, the intensity distributions are calculated by adding diffraction intensities under many slightly different illuminations with varying incident angles and transverse shifts. The diffraction patterns corresponding to the same illuminating position of Fig.\ref{Fig.2}(a) is shown in Fig.\ref{Fig.2}(b), where the intensity is obviously blurred. This coincides with Eq.(13). The reconstructed modulus and phase of the specimen with stable illumination are very clear as shown in Figs.\ref{Fig.2}(d) and (g), respectively. Figs.\ref{Fig.2}(e) and (h) show the reconstructed complex transmission of the specimen with the unstable illumination, where the resolution is apparently decreased compared with Figs.\ref{Fig.2}(d) and (g). Corrected diffraction patterns are obtained from the raw recorded data using the proposed method. A Gaussian function is used to slightly blur the recorded intensities with $K_v=1.24mm^{-1}$. And the obtained blurred diffraction patterns are subtracted from the raw recorded diffraction patterns. Then the modified intensity patterns are obtained by adding the amplified subtractions to the raw data with $\beta=2.5$. As shown in Fig.\ref{Fig.2}(c), the contrast of the corrected diffraction pattern is obviously improved compared with the raw data and the distribution is similar to that of Fig.\ref{Fig.2}(a). The reconstructed modulus and phase of the specimen using the corrected intensities are shown in Figs.\ref{Fig.2}(f) and (i), respectively, where the resolution is remarkably improved compared with Figs.\ref{Fig.2}(e) and (h). The insets in Figs.\ref{Fig.2}(d)-(f) and (g)-(i) show the reconstructed modulus and phase, respectively, of the illumination field, indicating that the quality of the retrieved illumination is also improved using the proposed method.To quantify the performance of the proposed method, the normalized root-mean-square error metric [13] is calculated, as shown in Fig. 3. The accuracy of the reconstruction are obviously improved when the proposed method is used to correct the recorded diffraction patterns before the iterative computation.\par
\begin{figure}[htbp]
\centering
\includegraphics[width=2.5in]{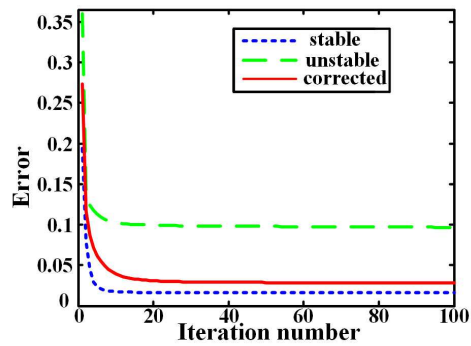}
\caption{Progress of the error of the conventional PIE method for a static imaging system (blue dashed line),an an unstable system (green dashed line); the proposed method for an unstable system(red line).}
\label{Fig.3}
\end{figure}
\subsection{Experimental result}
\begin{figure}[htbp]
\centering 
\includegraphics[width=4.5in]{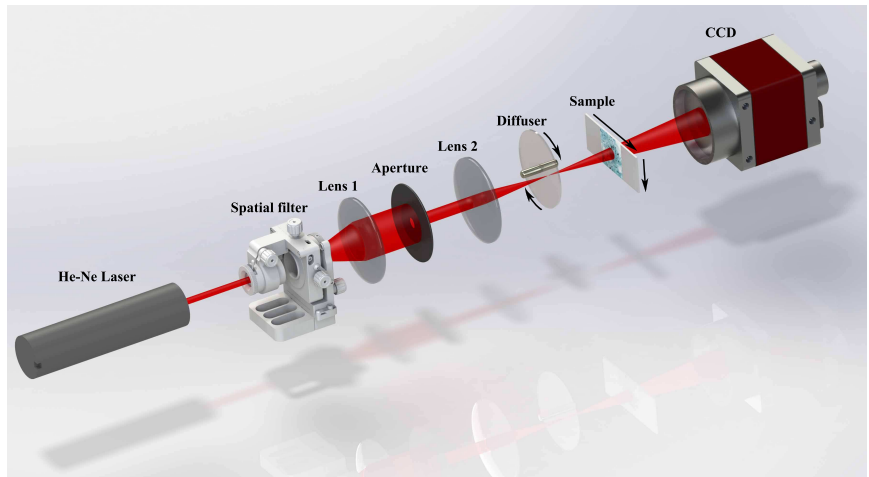}
\caption{Schema of the experimental setup.}
\label{Fig.4}
\end{figure}
The proof-of-principle experiment is conducted with visible light as shown in Fig. \ref{Fig.4}, where the divergent laser beam from a He-Ne laser is slightly diffused by a rotating plastic diffuser before irradiating on the sample to simulate the instability of the imaging system. A cross section of a monocotyledon placed on a 2D translation stage is used as the sample. The illumination area on the specimen is $\backsim$ 2.5 mm, and $10\times10$ diffraction patterns are recorded by the CCD camera while the sample is scanned relative to the instable beam with a step size of 0.37 mm.\par
\begin{figure}[htpb]
\centering
\includegraphics[width=3.5in]{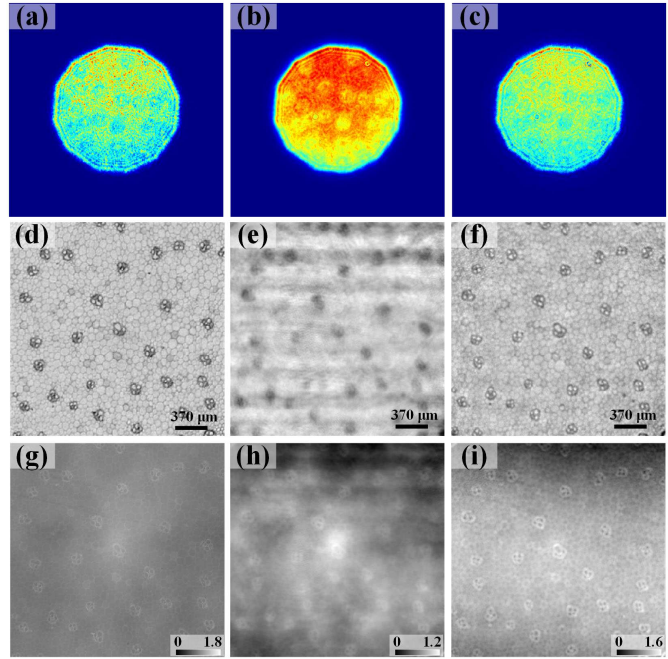}
\caption{Diffraction patterns (a) in the stable case, (b) in the unstable case, and (c) modified using the proposed method; reconstructed modulus of the specimen (d) in the stable case, (e) in the unstable case, and (f) obtained using the proposed method; reconstructed phase of the specimen (g) in the stable case, (h) in the unstable case, and (i) obtained using the proposed method.}
\label{Fig.5}
\end{figure}
Recorded intensities obtained with stable and unstable laser beams are shown in Figs.\ref{Fig.5}(a) and (b), respectively. It is clear that the recorded intensities using unstable imaging system are totally blurred. In order to overcome this limitation, the newly proposed method is applied to modify the recorded patterns. The Gaussian function with $K_v=0.9355mm^{-1}$ is adopted to slightly blur the recorded diffraction intensities. According to the proposed method, the corrected diffraction patterns are obtained with $\beta=5$. As shown in Fig.\ref{Fig.5}(c), the contrast of the corrected diffraction pattern is remarkably improved and similar to that shown in Fig.\ref{Fig.4}(a). Figs.\ref{Fig.5}(d) and (g) show the reconstructed modulus and phase, respectively, of the sample with stable illumination, where the fine structures of individual cells are clear. Figs.\ref{Fig.5}(e) and (h) show the reconstructed image of the specimen with the unstable imaging system, where the individual cell is hardly resolved. With the corrected diffraction patterns, the reconstructed image of the specimen shown in Figs.\ref{Fig.5}(f) and (i) are generated. Here, the quality of the reconstructed images is distinctly improved, and the images obtained are roughly identical to those for the stable illumination.\par
\begin{figure}
\centering
\includegraphics[width=4in]{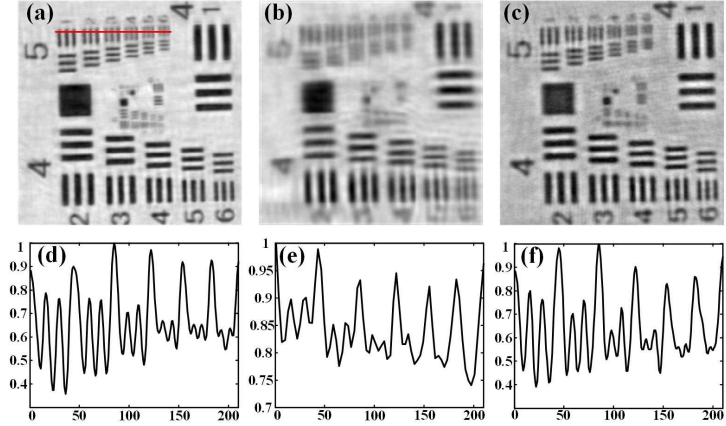}
\caption{Reconstructed modulus distribution of the USAF 1951 resolution target for (a) stable illumination, (b) unstable illumination, and (c) the proposed method; normalized value of the red line in (a) for (d) stable illumination, (e) unstable illumination, and (f) the proposed method.}
\label{Fig.6}
\end{figure}
To quantify the resolvability of the proposed method, a USAF1951 target is measured. As shown in Fig.\ref{Fig.6}, group 5, which is resolvable for the stable illumination, is obviously blurred for the unstable system. After the proposed method is applied, the elements in Group 5 become distinguished, as shown in Fig.\ref{Fig.6}(c) and (f).\par
\section{Discussion}
\begin{figure}[htbp]
\centering
\includegraphics[width=3in]{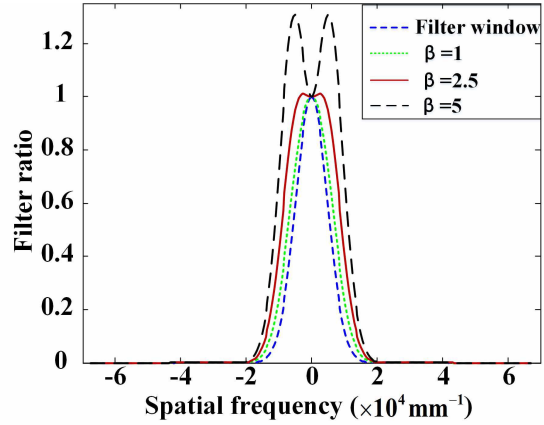}
\caption{Filter window obtained with unstable illumination and by modification with different values of $\beta$}
\label{Fig.7}
\end{figure}
\subsection{The influence of $\beta$}
To show the dependency of the resolution-improvement effect on the value of $\beta$, a numerical simulation is performed with the same parameters that were used for the simulation in Fig.\ref{Fig.2}. Fig.\ref{Fig.7} shows the effect of the proposed method with various curves, where the blue dashed line indicates the low-pass filter related to the vibration of the imaging system, and the other curves show the increased width of the low-pass filter with varying $\beta$. The width of the low-pass filter increases remarkably with increasing $\beta$, and for $\beta = 2.5$, the width of $[1+\beta-\beta exp(-C_v^2 u^2 )]exp(-C^2 u^2 )$ is about twice that of the $exp(-C^2u^2)$, and accordingly the resolution of the final reconstruction with these generated diffraction patterns can be remarkably improved. However, when $\beta = 5$, the low-pass filter has the shape of the black dashed line in Fig.\ref{Fig.7}; that is, the strength of the high-frequency components is over-amplified. Thus, the constant $\beta$ should be carefully selected to assure a satisfactory final reconstruction.\par
\begin{figure}[htbp]
\centering
\includegraphics[width=\linewidth]{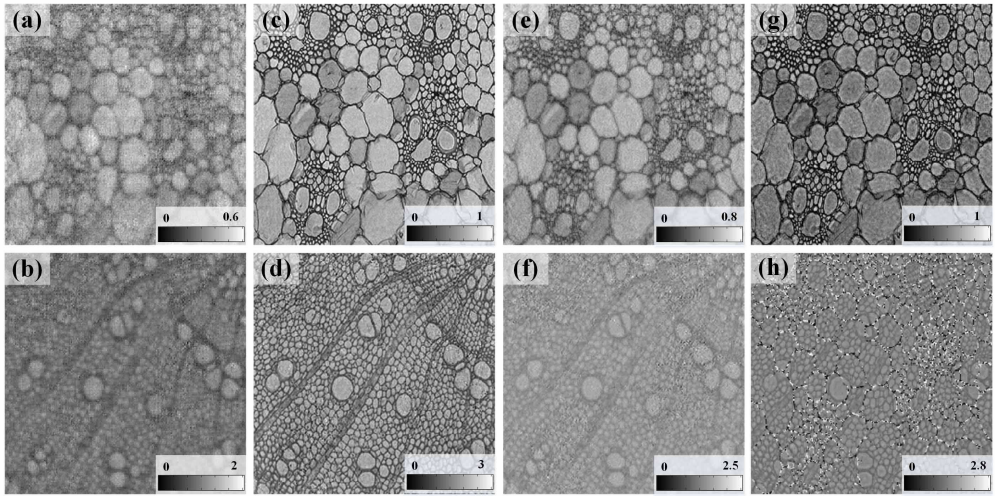}%
\caption{Reconstructed (a) modulus and (b) phase using unstable system; reconstructed (c) modulus and (d) phase using modified intensity patterns with $\beta=2.5$; reconstructed (e) modulus and (f) phase using modified intensity patterns with $\beta=1$; reconstructed (g) modulus and (h) phase using modified intensity patterns with $\beta=5$.}
\label{Fig.8}
\end{figure}

Fig.\ref{Fig.8} shows the reconstruction results with different $\beta$ values according to the aforementioned analysis, where Fig.\ref{Fig.8} (a) and (b) are the reconstructed modulus and phase images, respectively, obtained using the raw data acquired with the unstable system. Fig.\ref{Fig.8}(c) to (h) show the reconstructed modulus and phase images with $\beta = 2.5$, 1, and 5. These results indicate that the modulus and phase reconstructed using the proposed method produce a better reconstruction quality than those obtained directly with unstable data. On the other hand, the effect of the proposed method depends on the properly chosen value of $\beta$, which is coincident with the curves shown in Fig.\ref{Fig.7}.\par
\subsection{Robustness of the proposed method}
In the aforementioned theoretical analysis, the vibration of the illumination is assumed to have a normal distribution, and another Gaussian function is used to convolve with the recorded intensities to slightly blur the patterns. But the characteristics of a real imaging system are unknown and may be more complex, so it appears that the proposed method is difficult to implement for real experiments. However, the fundamental principle of the proposed method is to recover the high-frequency components lost because of the instability of the radiation source, and other functions that can realize this purpose can replace the Gaussian function in the analysis. For example, circular functions, triangular functions, and para-curves can be adopted to slightly blur the diffraction patterns for improving the resolution without considering the exact properties of the vibration of the imaging system. This feature makes the proposed method more applicable to real experiments and is demonstrated by the following simulations and experiments.\par
\begin{figure}[htbp]
\centering
\includegraphics[width=5in]{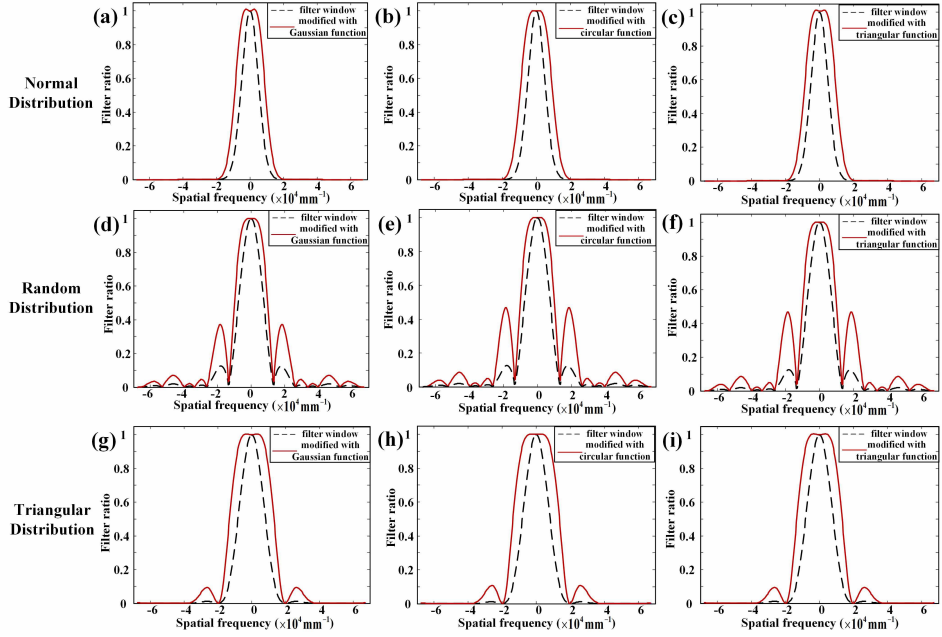}
\caption{When the vibration follows normal distribution, filter window (a) modified with Gaussian function ($K_v=1.24 mm^{-1}$, $\beta=2.5$); (b) modified with circular function ($K_v=1.1694 mm^{-1}$, $\beta=3.5$); (c) modified with triangular function ($K_v=0.8771mm^{-1}$, $\beta=4$); when the vibration follows random distribution, filter window (d) modified with Gaussian function ($K_v=0.8268 mm^{-1}$, $\beta=2.6$); (e) modified with circular function ($K_v=0.8771mm^{-1}$, $\beta=3.5$); (f) modified with triangular function ($K_v=0.7309 mm^{-1}$, $\beta=3.5$); when the distribution of the vibration follows triangular function , filter window (g) modified with Gaussian function ($K_v=0.8268 mm^{-1}$, $\beta=2$); (h) modified with circular function ($K_v=0.8771 mm^{-1}$, $\beta=3.2$); (i) modified with triangular function ($K_v=0.7309 mm^{-1}$, $\beta=2.4$).}
\label{Fig.9}
\end{figure}
When the vibration of the imaging system follows a normal distribution, the parameters are the same as that were used for the simulations in Fig.\ref{Fig.2}. As previously discussed, when the diffraction patterns are deliberately blurred via convolution with another Gaussian function $exp(-k^2/K_v^2)$, the width of the low-pass filter shown by the red line in Fig.\ref{Fig.9}(a) becomes much wider than that for the dashed line related to the instability. The same results can be obtained by using the circular function $circ(k/K_v)$ (Fig.\ref{Fig.9}(b)) and the triangular function $\Lambda (k/K_v )$  (Fig.\ref{Fig.9}(c)). For the case where the vibration of the imaging system follows a random distribution with the maximum extent of $52.5 \mu m$ in the detector plane, the low-pass filter window of the recorded intensity pattern is shown by the black dashed line in Fig.\ref{Fig.9}(d). The modified results obtained by the deliberately blurring diffraction patterns via convolution with a Gaussian function, circular function, and triangular function are shown as the red lines in Fig.\ref{Fig.9}(d) to (f), respectively. When the vibration of the imaging system has a triangular distribution and results in an extent of $36.2 \mu m$ in the detector plane, the low-pass filter window of the recorded intensity pattern has the profile of the black dashed line shown in Fig.\ref{Fig.9}(g). The modified results obtained by the deliberately blurring diffraction patterns via convolution with a Gaussian function, circular function, and triangular function are shown as the red lines in Fig.\ref{Fig.9}(g) to (i), respectively.\par
\begin{figure}[htbp]
\centering
\includegraphics[width=4in]{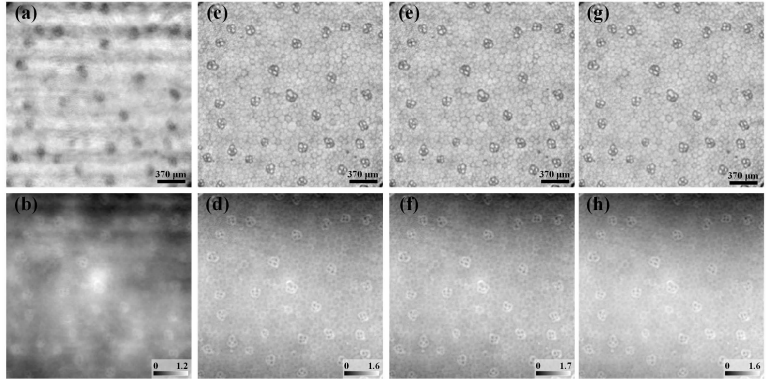}%
\caption{Reconstructed (a) modulus and (b) phase of the sample with unstable illumination; reconstructed (c) amplitude and (d) phase using the intensity patterns modified with Gaussian function; reconstructed (e) modulus and (f) phase using the intensity patterns modified with circular function; reconstructed (g) modulus and (h) phase using the intensity patterns modified with triangular function.}
\label{Fig.10}
\end{figure}

And an experiment is carried out to verify the practicality of the proposed method, and the parameters are the same with Fig.\ref{Fig.5}. The reconstructed images shown in  Fig.\ref{Fig.10}(a) and (b) are the reconstructions using the raw data recorded with instable illumination, which are seriously blurred. The reconstructed images shown in Fig.\ref{Fig.10} (c) and (d) are obtained by using the Gaussian function with $K_v=0.9355mm^{-1}$ and $\beta=5$ to do the resolution improvement with the proposed method. Fig.\ref{Fig.10} (e) and (f) are obtained by using intensities modified with circular function with a diameter $K_v=1.4033mm^{-1}$ and $\beta=8$ to do the reconstruction. Fig.\ref{Fig.10} (g) and (h) are obtained by using intensities modified with triangular function with a diameter $K_v=1.1694mm^{-1}$ and $\beta=6$ to do the reconstruction. We can find that their effects on the resolution improvement are almost the same. These results shows that the effect of the proposed method is independent of the vibrating model of the imaging system; thus, the method does not require prior knowledge of the vibrating properties.\par
\section{Conclusion}
In conclusion, a simple method is proposed to relax the requirement of ptychography for the stability of the imaging system. The influence of the instability of the imaging system can be described as a low-pass filter acting on the ideal diffraction intensities; the high-frequency components are lost in the recorded data, generating a blurred image in the final reconstruction. Using the proposed method, the recorded intensity patterns are corrected by subtracting the deliberately blurred diffraction patterns from the original ones and then adding the amplified subtracted patterns to the recorded data. The resolution of the final reconstruction can be significantly improved by using the corrected diffraction patterns. The feasibility of the proposed method is demonstrated via both numerical simulations and experiments on an optical bench. Because the proposed method does not require any prior knowledge of the characteristics of the illumination beam or massive calculation during the iterative process, it is an easy approach for acquiring a high-quality reconstruction with an unstable imaging system. The method can be extended to other CDI techniques for imaging with short-wavelength irradiation, such as free electrons or soft X-ray lasers.

\section*{Funding}
National Natural Science Foundation of China (No. 61675215).


\begin{thebibliography}{99}

\bibitem{1}
R.~W. Gerchberg, "A practical algorithm for the determination of phase from image and diffraction plane pictures," Optik {\bfseries 35}, 237 (1972).
\bibitem{2}
J.~R. Fienup, "Phase retrieval algorithms: a comparison," \ao {\bfseries 21}, 2758--2769 (1982).
\bibitem{3}
J.~Miao, P.~Charalambous, J.~Kirz, and D.~Sayre, "Extending the methodology of x-ray crystallography to allow imaging of micrometre-sized non-crystalline specimens," \nat {\bfseries 400}, 342--344 (1999).
\bibitem{4}
I.~K. Robinson, I.~A. Vartanyants, G.~Williams, M.~Pfeifer, and J.~Pitney, "Reconstruction of the shapes of gold nanocrystals using coherent x-ray diffraction," \prl {\bfseries 87}, 195505 (2001).
\bibitem{5}
D.~Shapiro, P.~Thibault, T.~Beetz, V.~Elser, M.~Howells, C.~Jacobsen, J.~Kirz, E.~Lima, H.~Miao, A.~M. Neiman , "Biological imaging by soft x-ray diffraction microscopy," Proceedings of the National Academy of Science {\bfseries 102}, 15343--15346 (2005).
\bibitem{6}
J.~Zuo, I.~Vartanyants, M.~Gao, R.~Zhang, and L.~Nagahara, "Atomic resolution imaging of a carbon nanotube from diffraction intensities," Science {\bfseries 300}, 1419--1421 (2003).
\bibitem{7}
J.~Miao, T.~Ishikawa, I.~K. Robinson, and M.~M. Murnane, "Beyond crystallography: Diffractive imaging using coherent x-ray light sources,"Science {\bfseries 348}, 530--535 (2015).
\bibitem{8}
P.~Bao, F.~Zhang, G.~Pedrini, and W.~Osten, "Phase retrieval using multiple illumination wavelengths," \ol {\bfseries 33}, 309--311(2008).
\bibitem{9}
V.~Y. Ivanov, M.~Vorontsov, and V.~Sivokon, "Phase retrieval from a set of intensity measurements: theory and experiment," \josaa {\bfseries 9}, 1515--1524 (1992).
\bibitem{10}
F.~Zhang and J.~Rodenburg, "Phase retrieval based on wave-front relay and modulation," \prb {\bfseries 82}, 121104 (2010).
\bibitem{11}
H.~Tao, S.~P. Veetil, X.~Pan, C.~Liu, and J.~Zhu, "Lens-free coherent modulation imaging with collimated illumination," Chinese Optics Letters {\bfseries 14}, 071203 (2016).
\bibitem{12}
H.~Faulkner and J.~Rodenburg, "Movable aperture lensless transmission microscopy: a novel phase retrieval algorithm," \prl {\bfseries 93}, 023903 (2004).
\bibitem{13}
A.~M. Maiden and J.~M. Rodenburg, "An improved ptychographical phase retrieval algorithm for diffractive imaging," Ultramicroscopy {\bfseries 109},1256--1262 (2009).
\bibitem{14}
H.~Faulkner and J.~Rodenburg, "Error tolerance of an iterative phase retrieval algorithm for moveable illumination microscopy," Ultramicroscopy {\bfseries 103}, 153--164 (2005).
\bibitem{15}
K.~Stachnik, I.~Mohacsi, I.~Vartiainen, N.~Stuebe, J.~Meyer, M.~Warmer, C.~David, and A.~Meents, "Influence of finite spatial coherence on ptychographic reconstruction," \apl {\bfseries 107}, 011105 (2015).
\bibitem{16}
N.~Burdet, X.~Shi, D.~Parks, J.~N. Clark, X.~Huang, S.~D. Kevan, and I.~K. Robinson, "Evaluation of partial coherence correction in x-ray ptychography," \ol {\bfseries 3}, 5452--5467 (2015).
\bibitem{17}
L.~Whitehead, G.~Williams, H.~Quiney, D.~Vine, R.~Dilanian, S.~Flewett,  K.~Nugent, A.~G. Peele, E.~Balaur, and I.~McNulty, "Diffractive imaging using partially coherent x rays,"\prl {\bfseries 103}, 243902 (2009).
\bibitem{18}
J.~N. Clark and A.~G. Peele, "Simultaneous sample and spatial coherence characterisation using diffractive imaging," \apl {\bfseries 99}, 154103 (2011).
\bibitem{19}
B.~Chen, R.~A. Dilanian, S.~Teichmann, B.~Abbey, A.~G. Peele, G.~J. Williams, P.~Hannaford, L.~Van~Dao, H.~M. Quiney, and K.~A. Nugent, "Multiple wavelength diffractive imaging," \pra {\bfseries 79}, 023809 (2009).
\bibitem{20}
P.~Thibault and A.~Menzel, "Reconstructing state mixtures from diffraction measurements," \nat {\bfseries 494}, 68--71 (2013).
\bibitem{21}
D.~J. Batey, D.~Claus, and J.~M. Rodenburg, "Information multiplexing in ptychography," Ultramicroscopy {\bfseries 138}, 13--21 (2014).
\bibitem{22}
W.~Yu, S.~Wang, S.~Veetil, S.~Gao, C.~Liu, and J.~Zhu, "High-quality image reconstruction method for ptychography with partially coherent illumination," \prb {\bfseries 93}, 241105 (2016).
\bibitem{23}
S.~Dong, P.~Nanda, K.~Guo, J.~Liao, and G.~Zheng, "Incoherent fourier ptychographic photography using structured light," Photonics Research {\bfseries 3}, 19--23 (2015).
\bibitem{24}
K.~Guo, S.~Dong, and G.~Zheng, "Fourier ptychography for brightfield, phase, darkfield, reflective, multi-slice, and fluorescence imaging," IEEE Journal of Selected Topics in Quantum Electronics {\bfseries 22}, 77--88 (2016).
\bibitem{25}
F.~Wei, J.-Y. Choi, and S.~Rah, "Experiences of the long term stability at sls," Proc. AIP  {\bfseries 879}, 38--41(2007).
\bibitem{26}
V.~Schlott, M.~Boge, B.~Keil, P.~Pollet, and T.~Schilcher, "Fast orbit feedback and beam stability at the Swiss Light Source," Proc. AIP {\bfseries 732}, 174--181{2004}.
\bibitem{27}
F.~Zhang, I.~Peterson, J.~Vila-Comamala, A.~Diaz, F.~Berenguer, R.~Bean, B.~Chen, A.~Menzel, I.~K. Robinson, and J.~M. Rodenburg, "Translation position determination in ptychographic coherent diffraction imaging," \opex {\bfseries 21}, 13592--13606 (2013).
\bibitem{28}
A.~Maiden, M.~Humphry, M.~Sarahan, B.~Kraus, and J.~Rodenburg, "An annealing algorithm to correct positioning errors in ptychography," Ultramicroscopy {\bfseries 120}, 64--72 (2012).
\bibitem{29}
M.~Guizar-Sicairos and J.~R. Fienup, "Phase retrieval with transverse translation diversity: a nonlinear optimization approach," \opex {\bfseries 16}, 7264--7278 (2008).
\bibitem{30}
M.~Odstrcil, P.~Baksh, S.~Boden, R.~Card, J.~Chad, J.~Frey, and W.~Brocklesby, "Ptychographic coherent diffractive imaging with orthogonal probe relaxation," \opex {\bfseries 24}, 8360--8369 (2016).
\bibitem{31}
J.~N. Clark, X.~Huang, R.~J. Harder, and I.~K. Robinson, "Dynamic imaging using ptychography," \prl {\bfseries 112}, 113901 (2014).
  
\end{thebibliography}
\end{document}